\documentstyle[prl,aps,twocolumn]{revtex}
\begin{document}
\voffset 0.5in
\draft
\wideabs{
\title{Density-functional study of the Mott gap in the Hubbard model}
\author{N.~A. Lima and L.~N. Oliveira}
\address{Departamento de F\'{\i}sica e Inform\'atica,
Instituto de F\'{\i}sica de S\~ao Carlos,
Universidade de S\~ao Paulo,\\
Caixa Postal 369, 13560-970 S\~ao Carlos, SP, Brazil}
\author{K. Capelle}
\address{Departamento de Qu\'{\i}mica e F\'{\i}sica Molecular,
Instituto de Qu\'{\i}mica de S\~ao Carlos,
Universidade de S\~ao Paulo,\\
Caixa Postal 780, S\~ao Carlos, 13560-970 SP,
Brazil \\ \hspace*{1cm}\\
{\bf accepted by Europhysics Letters (2002)\\}}
\date{\today}
\maketitle
\begin{abstract}
We study the Mott insulating phase of the one-dimensional Hubbard model using
a local-density approximation (LDA) that is based on the Bethe Ansatz (BA). 
Unlike conventional functionals the BA-LDA has an explicit derivative
discontinuity. We demonstrate that as a consequence of this discontinuity
the BA-LDA yields the correct Mott gap, independently of the strength of
the correlations. A convenient analytical formula for the Mott gap in the
thermodynamic limit is also derived. We find that in one-dimensional quantum 
systems the contribution of the discontinuity to the full gap is more important
than that of the band-structure gap, and discuss some consequences this
finding has for electronic-structure calculations.  
\end{abstract}


}
\newcommand{\be}{\begin{equation}}
\newcommand{\ee}{\end{equation}}
\newcommand{\bea}{\begin{eqnarray}}
\newcommand{\eea}{\end{eqnarray}}
\newcommand{\bi}{\bibitem}

\newcommand{\ep}{\epsilon}
\newcommand{\s}{\sigma}
\newcommand{\p}{{\bf \pi}}
\newcommand{\r}{({\bf r})}
\newcommand{\rp}{({\bf r'})}
\newcommand{\rpp}{({\bf r''})}

\newcommand{\ua}{\uparrow}
\newcommand{\da}{\downarrow}
\newcommand{\la}{\langle}
\newcommand{\ra}{\rangle}
\newcommand{\dg}{\dagger}

The intricacies of the correlation-induced metal-insulator transition
(`Mott transition') have fascinated and challenged physicists for many
years, and a multitude of methods has been brought to bear on the problem
\cite{mott}. Here we study the Mott insulator from the point of view of 
density-functional theory (DFT), a method whose full potential for exploring 
the Mott phenomenon has not been widely recognized. It may be particularly
timely to study the prospects for a DFT treatment of the Mott insulating
phase because this is one of the two main phases of the one-dimensional
Hubbard model (1DHM), the paradigmatic model for low-dimensional 
quantum systems. The recent interest in nanotechnology has brought
quasi one-dimensional systems, such as quantum wires and carbon 
nanotubes, into the focus of current research in many-body physics
and materials science. DFT, on the other hand, is the {\it de facto}
standard method for the calculation of the electronic structure of materials,
but its applicability to the very peculiar phases typical of 
low-dimensional quantum systems has not been systematically investigated.
Here we report on the application of a very recently developed density 
functional \cite{luttlett} to the Mott insulating phase of the 1DHM.

DFT is often applied in a single-particle mode, in which the eigenvalues of 
the Kohn-Sham (KS) equation are interpreted as a mean-field 
approximation to the quasi-particle energies. It is well known
that this approximation breaks down for systems with strong correlations,
of which the Mott insulator is an example. On the other hand, as long as one
restricts oneself to working with the particle density, the total
ground-state energy, and quantities derivable from these, DFT is a
rigorous many-body theory, whose predictive power is only limited by the
quality of available approximations for the exchange-correlation ($xc$)
functional. It is in this many-body mode that we employ DFT here. 

The choice of the appropriate approximation for the $xc$ functional depends on 
the system under study. Here we investigate the 1DHM, since this model 
represents a particularly clear case of a Mott transition, free of unessential 
complications: for nonzero interaction strength ($U>0$) the 1DHM is a Mott 
insulator at half filling, whereas it is a Luttinger liquid (i.e., a 
one-dimensional metal) at all other fillings \cite{liebwu,voit,schlottman}. 
Following pioneering work of Gunnarsson, Sch\"onhammer, and Noack \cite{gs,sgn}
we have recently constructed an approximate $xc$ functional for this model
\cite{luttlett}. This functional is a local-density approximation (LDA) that
is based on the exact Bethe Ansatz (BA) solution of the homogeneous 1DHM   
in the thermodynamic limit. For $N\leq 1$ it takes the form
\be
E^{BALDA}_{xc}[n_i,U]=\sum_i e^<_{xc}(n_i,U),
\label{ldaform}
\ee
where $e^<_{xc}(n,U)=e^<(n,U)-e^<(n,0)-Un^2/4$ is the  per-site $xc$ energy of
the homogeneous infinite 1DHM, and $e^<(n,U)$ is the corresponding total 
energy, for which we have given an accurate parametrization, based on the 
Bethe Ansatz, in Ref.~\cite{luttlett}. 
For $n_i>1$ a particle-hole transformation shows that 
$e^<_{xc}(n,U)$ must be replaced by $e_{xc}^>(n,U)=e_{xc}^<(2-n,U)$. 
Here we apply this functional to the Mott insulator.

Our starting point is the exact expression for the energy gap of a
many-body system, $\Delta = I-A$, where
$ I=E(N-1)-E(N)$
is the system's ionization energy, 
$A=E(N)-E(N+1)$
its electron affinity, and $E(N)$ the ground-state energy of the $N$-particle
system. Hence,
\be
\Delta=E(N+1)+E(N-1)-2 E(N).
\label{rep1}
\ee
In ensemble DFT one can derive an alternative, but equivalent, representation 
for the true energy gap, namely \cite{dftbook}
\be
\Delta=\ep_{k_{max}+1}(N)-\ep_{k_{max}}(N) + \Delta_{xc}
=:\Delta_{KS}+\Delta_{xc},
\label{rep2}
\ee
where $\ep_k(N)$ is the $k$'th single-particle energy (KS eigenvalue)
of the $N$-particle system, and $\Delta_{xc}$ is the discontinuity of the
exchange-correlation potential\cite{lp}
\be
\Delta_{xc}=\left.\frac{\delta E_{xc}[n]}{\delta n\r}\right|_{N+\delta}
-\left.\frac{\delta E_{xc}[n]}{\delta n\r}\right|_{N-\delta},
\ee
where $\delta \to 0^+$ is an infinitesimal change in the total particle 
number $N$. According to ensemble DFT, for open systems the total energy
for fractional particle number is given by a straight-line interpolation
between the energies obtained for integer particle number \cite{lp}. 
It is this fact that guarantees the equivalence of both representations
\cite{dftbook}.

For the homogeneous infinite 1DHM at $n=1$ one can evaluate both
Eq.~(\ref{rep1}) and Eq.~(\ref{rep2}) analytically from the parametrization 
underlying the BA-LDA. In either case we find
\be
\Delta_{hom}^{BALDA}(U) = U+4\cos\left(\frac{\pi}{\beta(U)}\right),
\label{homgap}
\ee
where $\beta(U)$ is calculated from \cite{luttlett}
\be
-\frac{2\beta}{\pi} \sin\left(\frac{\pi}{\beta}\right) =
-4\int _0^\infty dx
\frac{J_0(x)J_1(x)}{x[1+\exp(Ux/2)]},
\label{betadet}
\ee
and $J_0$ and $J_1$ are zero and first order Bessel functions.
(Here and below we take $t=1$ as our unit of energy.) 

Representation (\ref{rep2}) shows that there are in general two contributions
to the many-body gap, one being the KS gap, $\Delta_{KS}$ (this is the
contribution obtained in a standard band-structure calculation), the other
the correlation-induced Mott gap, $\Delta_{xc}$, which is a pure
many-body effect. Unfortunately, the most commonly employed approximate 
$xc$ functional, the conventional {\it ab initio} LDA, does not have a
discontinuity at integer $N$, so that $\Delta_{xc}^{LDA}=0$ and
$\Delta^{LDA}=\Delta^{LDA}_{KS}$. This is the origin of the well known
band-gap problem of the LDA \cite{lp}. 
The same applies to common gradient-corrected functionals. 
In the case of {\it ab initio} calculations one finds it thus 
necessary to reach beyond DFT and employ more or less systematically derived
correction terms, such as the scissors operator \cite{scissors}, LDA+U
\cite{ldaplusu}, or LDA+SIC \cite{sic}.
However, the situation is different for the BA-LDA. Due to the particle-hole 
transformation involved in its construction this functional naturally has a 
discontinuity at half filling (occupation number $n=1$), where the underlying 
homogeneous infinite 1DHM undergoes its metal-insulator transition.
The BA-LDA is thus uniquely suited to test the density-functional theory
of the many-body energy gap. 

To begin our analysis we present, in the left half of Fig.~\ref{fig1}, the 
energy gap of a homogeneous 1DHM with $L=10$ sites, calculated with four 
different approaches:
(i) BA-LDA gap calculated numerically from representation (\ref{rep1}), 
(ii) energy gap obtained from representation (\ref{rep1}) if the
pseudo LDA of Refs.~\cite{gs,sg} (GS-LDA) is used instead of the BA-LDA 
(note that that functional does not have a discontinuity),
(iii) BA-LDA gap calculated analytically for an infinite homogeneous 
system from Eq.~(\ref{homgap}), and
(iv) the exact gap, obtained by numerical diagonalization.
Clearly, the continuous functional drastically underestimates the exact gap.
In the right half of Fig.~\ref{fig1} we present the same results for a 150-site 
system. The last column is now missing since a system of this size cannot
be treated anymore by numerical exact diagonalization. 
However, precisely because $L$ is large, we now observe that the gap 
obtained numerically from the BA-LDA (the first bar) is close to the
analytical result of Eq.~(\ref{homgap}) (third bar).

The main result that follows from the data presented in Fig.~\ref{fig1} 
is that for $U=6$ the dominating contribution to the many-body 
gap is due to the discontinuity. We have repeated the same analysis
also for other values of $U$ and $L$, and found that the critical value
of $U$ for which the discontinuity gap $\Delta_{xc}$ becomes more important 
than the band-structure gap $\Delta_{KS}$ decreases rapidly with increasing 
$L$, e.g. $U_c(L=6)=5.2$ and $U_c(L=10)=4.2$, while $U_c(L=500)=1.8$.
For large systems and realistic values of $U$ the discontinuity is thus
the dominating contribution to the full gap. This observation is in agreement
with that of Refs.~\cite{sgn} and \cite{godby}, but contradicts the one of 
Ref.~\cite{sg}.

From our results it follows that the discontinuity contribution is absolutely 
crucial for obtaining the correct many-body gap from DFT. This implies that 
{\it in order to solve the band gap problem encountered in the DFT description 
of semiconductors and insulators it is not useful to  optimize a
continuous functional (e.g. by passing from LDA to GGA) or to try to
improve on the band-structure part of the calculation (which will only result
in an improved value for $\Delta_{KS}$); rather one should search for a
properly discontinuous functional.}
Interestingly, Ref.~\cite{harrison} reports quantitative agreement of the
KS gap calculated from the B3LYP GGA functional and the experimental gap for
a variety of semiconductors and insulators. These authors themselves conclude 
that this agreement is probably fortuitious --- an interpretation that is
confirmed by the present analysis.

Fig.~\ref{fig2} displays, for a 150-site system, the BA-LDA \cite{luttlett} 
gap and the pseudo-LDA \cite{gs,sg} gap, 
both calculated numerically from representation (\ref{rep1}) as
a function of the interaction $U$. The two full curves are asymptotic results
(valid in the thermodynamic limit) that can be extracted analytically
from the Bethe Ansatz solution of the 1DHM in the limits $U\to \infty$ and 
$U\to 0$, respectively. For $U\to \infty$ \cite{schlottman,ovchinnikov}
\be
\Delta(U\to \infty) = U-4 + (8 \ln 2)/U.
\label{asym}
\ee 
The leading-order term in this expansion, $U-4$, is also obtained immediately
from our Eq.~(\ref{homgap}), since for $U\to\infty$ one has $\beta=1$
\cite{luttlett}. For $U\to 0$, on the other hand \cite{schlottman,ovchinnikov},
\be
\Delta(U\to 0) = \frac{8}{\pi}\sqrt{U} \exp{\left(-\frac{2\pi}{U}\right)}.
\label{deltazero}
\ee
For $U=0$ Eq.~(\ref{deltazero}) goes to zero. This, too, is reproduced by
Eq.~(\ref{homgap}), since for $U=0$ one has $\beta=2$ \cite{luttlett}.
However, the non-analyticity of the exact expression (\ref{deltazero})
is not properly accounted for by Eq.~(\ref{homgap}). This is not a defect
of the LDA (which becomes exact in the thermodynamic limit of a homogeneous 
system), but of our parametrization of it, which does not incorporate
this nonanalyticity. Work on improved parametrizations is in progress. 

As clearly seen in Fig.~\ref{fig2}, the BA-LDA reproduces the limiting 
behaviours quite well, both for large and for small $U$, in spite of
the intrinsic error of the underlying parametrization. The pseudo LDA,
on the other hand, is incapable of reproducing the exact gap for
large $U$.

In Fig.~\ref{fig3} we study the evolution of the various contributions to the 
many-body gap as an originally infinite homogeneous system with $U=6$ is
gradually made finite. Both the many-body gap and the single-particle
gap scale approximately linearly with inverse system size.
At $1/L=0$, i.e., in the thermodynamic limit, the single-particle gap is
zero. A band-structure calculation would predict the system to be a metal
in this limit. The $xc$ contribution to the gap, however, is nonzero and
can be calculated from Eq.~(\ref{homgap}). For finite system sizes 
($1/L > 0$) the boundaries give rise to quantization of the single-particle 
states and a discrete set of energy levels. The KS gap then becomes nonzero.
The many-body gap is in this case calculated numerically from Eq.~(\ref{rep1}),
and the $xc$ gap is extracted from that result by means of Eq.~(\ref{rep2}).
As seen in the inset of Fig.~\ref{fig3}, the $xc$ gap is a nearly 
quadratic function of $1/L$, i.e., does not scale with the same power as
the single-particle gap.

Interestingly, the $xc$ gap in the 1DHM is larger than the single-particle gap 
all the way from $L \to \infty$ down to $L\sim O(1)$. Applied to realistic 
quasi one-dimensional systems, such as quantum wires and nanotubes, this 
observation implies that the $xc$ gap is of
paramount importance for nanophysics and device technology. Unfortunately,
this contribution is completely missed by standard band-structure
approaches, which only calculate the single-particle gap. 
Notable exceptions are the optimized-effective potential method
\cite{kli,kotani,grabogross} and the `electron-gas with a gap' approach
of Refs.~\cite{krieger,savin,mgga}. In view of our findings such methods 
may aquire additional significance for first-principles calculations of 
the electronic structure of low-dimensional systems.

In summary, we have demonstrated the applicability of the BA-LDA approach to
the Mott insulating phase of the 1DHM. We find that the discontinuity of the 
exchange-correlation functional makes a crucial contribution to the full gap, 
both for finite and for infinite systems, and for a wide range of values of 
$U$. A continuous functional drastically underestimates the exact gap, 
whereas the BA-LDA yields quantitatively reliable results with very modest 
computational effort. This establishes the BA-LDA, which was originally 
designed for the Luttinger-liquid phase \cite{luttlett}, as a computational 
tool also in the insulating phase of the 1DHM. 
Although the BA-LDA was constructed specifically for the 1DHM and cannot be 
used directly with the {\it ab initio} Hamiltonian, we believe our findings 
concerning the discontinuity to be robust and qualitatively valid also for 
{\it ab initio} calculations.
In particular, we expect that also for {\it ab initio} calculations a 
reliable determination of the electronic structure of quantum wires and 
nanotubes can only be achieved if a functional with a proper discontinuity 
is employed.

{\bf Acknowledgments}
This work was sup\-por\-ted by FAPESP and CAPES. We thank M.~F.~da Silva
for many discussions and providing us with the exact diagonalization
program. KC thanks E.~K.~U. Gross and A. Schindlmayr for helpful discussions.

\begin{figure}
\caption{Many-body energy gap of the one-dimensional Hubbard model for $U=6$.
Left: $L=10$ sites.
First bar: calculated from the BA-LDA using Eq.~(\ref{rep1}). 
Second bar: calculated from the continuous pseudo LDA, again using
Eq.~(\ref{rep1}).
Third bar: calculated from the analytical BA-LDA expression
(\ref{homgap}), valid in the thermodynamic limit $L\to \infty$. 
Fourth bar: exact result.
Right: same as left, but with the first two bars now calculated for
a system with $L=150$ sites. As a consequence of the system size the bar
with the exact result is missing.}
\label{fig1}
\end{figure}

\begin{figure}
\caption{Energy gap calculated from Eq.~(\ref{rep1}) versus interaction 
$U$ for a homogeneous system with $L=150$ sites. Squares: BA-LDA,
circles: pseudo LDA. The full and dashed curves are the asymptotically valid 
analytical results for $U\to \infty$ and $U\to 0$ of Eqs.~(\ref{asym}) and
(\ref{deltazero}), respectively.}
\label{fig2}
\end{figure}

\begin{figure}
\caption{Energy gap of a finite 1DHM as a function of inverse system size
$1/L$. Upper curve (triangles): many-body gap $\Delta$, calculated from 
Eq.~(\ref{rep1}). Lower curve (circles): Kohn-Sham gap $\Delta_{KS}$, 
calculated as a difference of single-particle eigenvalues.
Middle curve (squares): $xc$ gap $\Delta_{xc}$, calculated as the difference
between the two other curves [i.e., from Eq.~(\ref{rep2})].
Inset: $xc$ gap on a smaller scale, showing approximately parabolic behaviour
of $\Delta_{xc}$ as a function of $1/L$.}
\label{fig3}
\end{figure}

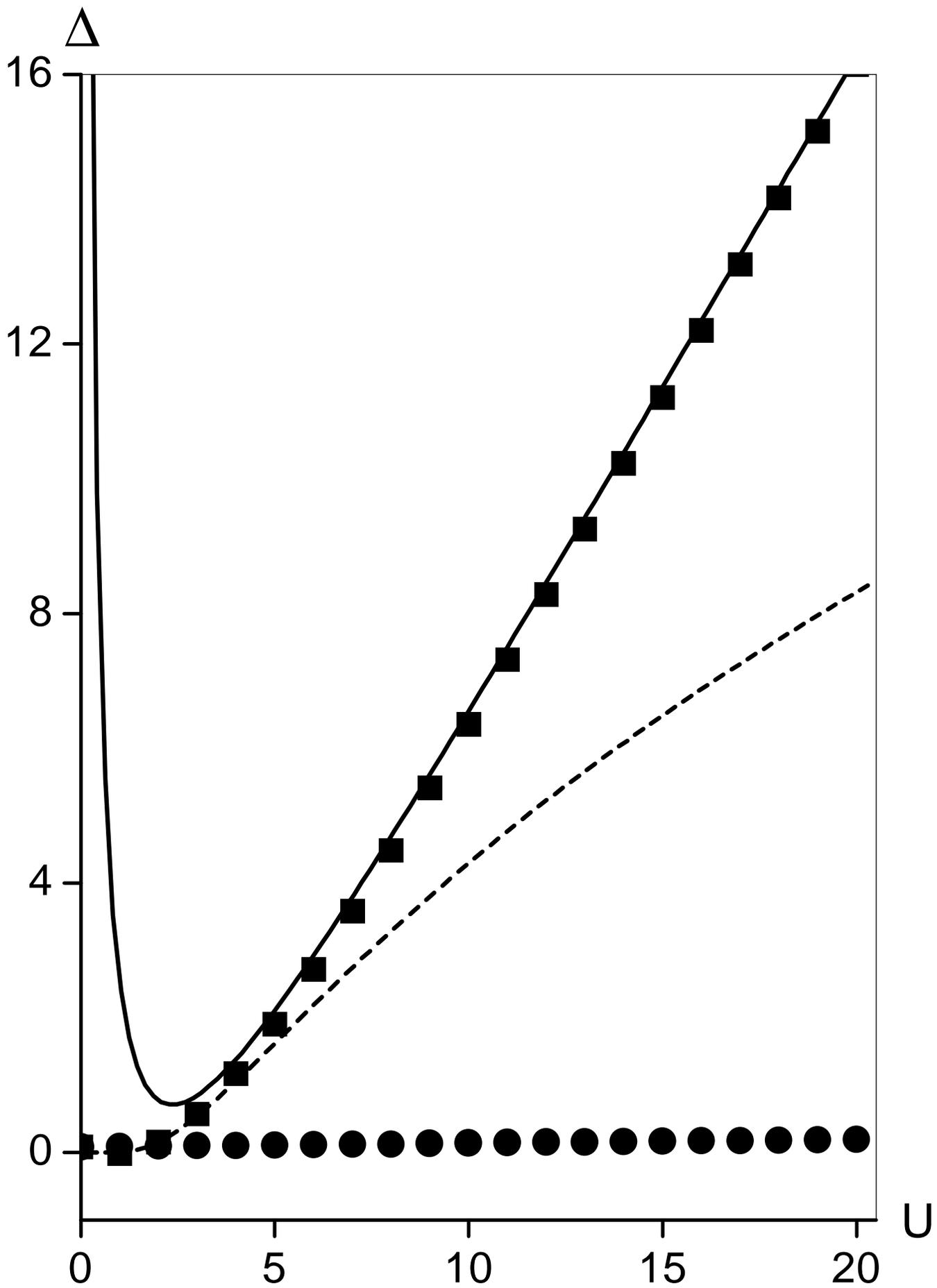
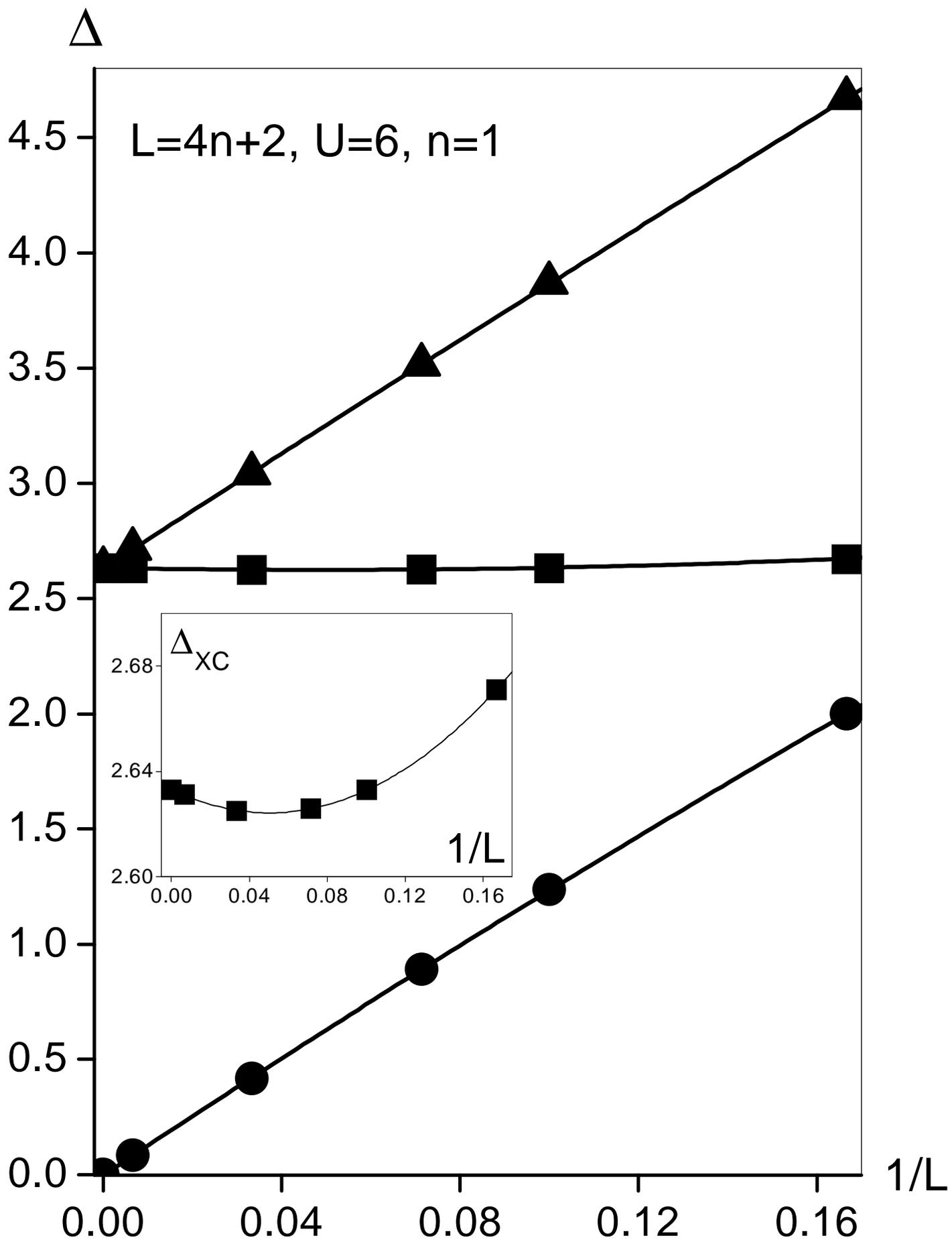
\end{document}